\ifdefined\pdfminorversion\pdfminorversion=7\fi
\PassOptionsToClass{balance=false}{acmart}
\documentclass[sigconf]{acmart}

\setcopyright{none}
\acmConference[WWW '27]{The ACM Web Conference 2027}{May 10--14, 2027}{Dublin, Ireland}
\acmYear{2027}
\copyrightyear{2027}
\acmDOI{}
\acmISBN{}
\usepackage{booktabs}
\usepackage{tabularx}
\usepackage{colortbl}
\definecolor{RQ4RowOne}{HTML}{FFC6C6}
\definecolor{RQ4RowTwo}{HTML}{FFE0E0}
\definecolor{RQ4RowThree}{HTML}{FFF2EE}
\graphicspath{{figures/}}
\usepackage{multirow}
\usepackage{placeins}
\usepackage{stfloats}
\usepackage{flafter}
\usepackage{enumitem}
\newcounter{weaveralgorithm}
\renewcommand{\theweaveralgorithm}{\arabic{weaveralgorithm}}
\newenvironment{weaveralgorithm}[2]{%
  \par\addvspace{7pt}\noindent
  \begin{minipage}{\linewidth}
  \refstepcounter{weaveralgorithm}\label{#2}%
  \fontsize{8.5}{10}\selectfont
  \hrule height 0.7pt\vspace{3pt}
  {\bfseries Algorithm \theweaveralgorithm\enspace #1\par}
  \vspace{3pt}\hrule height 0.35pt\vspace{4pt}
}{%
  \par\vspace{3pt}\hrule height 0.5pt
  \end{minipage}\par\addvspace{7pt}
}
\newcommand{\alginput}[1]{\noindent\textbf{Input:} #1\par}
\newcommand{\algoutput}[1]{\noindent\textbf{Output:} #1\par\vspace{3pt}}
\newenvironment{weaversteps}{%
  \begin{enumerate}[label=\arabic*:,leftmargin=1.6em,labelsep=0.35em,
    itemsep=1.3pt,parsep=0pt,topsep=1pt,partopsep=0pt]
}{\end{enumerate}}

\usepackage{tcolorbox}
\tcbuselibrary{skins,breakable}
\definecolor{PromptBlueBack}{HTML}{EEF4F8}
\definecolor{PromptBlueRule}{HTML}{41667D}
\definecolor{PromptGreenBack}{HTML}{F0F3EC}
\definecolor{PromptGreenRule}{HTML}{617C40}
\definecolor{PromptCreamBack}{HTML}{FCF0E6}
\definecolor{PromptCreamRule}{HTML}{D87921}
\newtcolorbox{weaverprompt}[3]{%
  enhanced,
  breakable,
  lines before break=5,
  colback=#1,
  coltext=black,
  frame hidden,
  boxrule=0pt,
  sharp corners,
  borderline west={2.4pt}{0pt}{#2},
  boxsep=0pt,
  left=8pt,right=7pt,top=7pt,bottom=7pt,
  before skip=7pt,after skip=7pt,
  fontupper=\fontsize{9}{11}\selectfont\raggedright,
  before upper={\setlength{\parindent}{0pt}\textbf{#3}\par\smallskip},
  parbox=false
}

\begin{document}

\title[Weaver: Incremental Preference Reasoning for Personalized RAG]{Weaver: Weaving Preference Evidence through Incremental Reasoning for Personalized Retrieval-Augmented Generation}
\author{Linyao Zheng}
\affiliation{%
  \institution{Xi'an Jiaotong University}
  \city{Xi'an}
  \country{China}
}

\author{Zhifang Mao}
\affiliation{%
  \institution{Beijing University of Posts and Telecommunications}
  \city{Beijing}
  \country{China}
}

\author{Xuhang Shi}
\affiliation{%
  \institution{Beijing University of Posts and Telecommunications}
  \city{Beijing}
  \country{China}
}

\begin{abstract}
Personalized retrieval-augmented generation requires extracting and applying user preferences from extended interaction histories. This task is challenging because preference evidence is scattered across past conversations and remains bound to specific temporal and situational contexts. Consequently, isolated fragments rarely provide complete support for a response, while preferences expressed under different conditions are easily conflated. To address these limitations, we propose Weaver, a framework that incrementally weaves memory-conditioned retrieval trajectories with compositional preference reasoning for personalized responses. Weaver operates on a Dual-Layer Retrieval and Reasoning Graph that couples an upper Intent Dependency Graph with a lower Memory-Conditioned Retrieval Trajectory Graph. Dependency Information Gap Tracking represents the information needed to answer the current request and its prerequisite dependencies in the upper graph. Memory-Conditioned Incremental Retrieval uses established premises and clues from retrieved conversations to locate missing evidence, recording the resulting retrieval trajectories in the lower graph. Scope-Aware Compositional Reasoning jointly interprets the accumulated evidence to determine which preferences apply to the request. It identifies complementary and alternative support and distinguishes conflicting claims within the same temporal and situational context. These conclusions iteratively update task premises and remaining information gaps, guiding subsequent retrieval to establish a coherent contextual basis for the final response. Experiments on personalized RAG benchmarks, particularly PersonaMem, demonstrate that Weaver outperforms leading baselines in both evidence completeness and answer quality. Specifically, Weaver achieves accuracies of 86.42\% on PersonaMem and 91.10\% on PrefEval, exceeding the strongest evaluated baselines by 16.01 and 7.56 percentage points, respectively.

\end{abstract}


\maketitle

\section{Introduction}
\label{sec:introduction}

Personalization enables Web systems to adapt information and services to users' current needs~\citep{brusilovsky2001adaptive}. Large language models extend this capability to generating responses tailored to individual users~\citep{baek2024klamp,puma2025,kumar2024longlamp}. Interaction histories provide preference evidence that is often absent from the current request~\citep{du-etal-2024-perltqa}. LLM-based Web agents use this evidence to interpret instructions and generate personalized responses~\citep{chen2024personalization}. As histories grow, conditioning on the full history introduces irrelevant context and makes preference following more difficult~\citep{prefeval2025,liu-etal-2024-lost}. Personalized retrieval-augmented generation (RAG) addresses this problem by selecting relevant historical records to inform the current response~\citep{salemi2024lamp,salemi2024optimization,wang2024emg,yu2026memweaver,zhang2026rfmem,amirizaniani2026pr2}. Selective access focuses the model on potentially useful preference evidence~\citep{richardson2023summarization,mysore-etal-2024-pearl}. However, relevance alone does not establish that the retrieved records sufficiently support a preference applicable to the current request~\citep{joren2025sufficient,tan-etal-2025-personabench}.

\begin{figure}[!t]
  \centering
  \includegraphics[width=\linewidth]{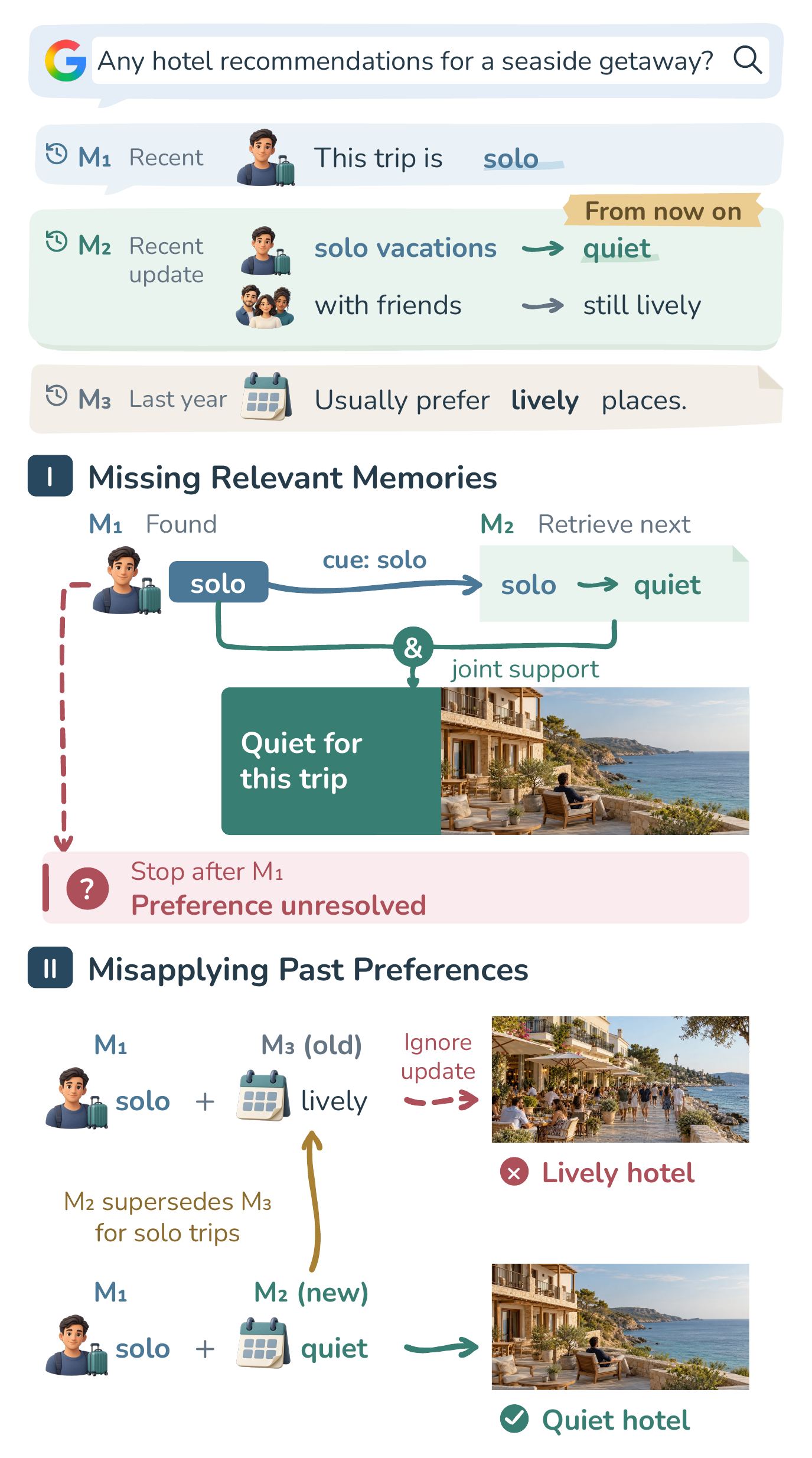}
  \caption{Two difficulties in preference inference. \textbf{I) Missing Relevant Memories.} Stopping at $M_1$ leaves the preference unresolved, while retrieving $M_2$ completes the support. \textbf{II) Misapplying Past Preferences.} Applying the older $M_3$ ignores the solo-trip update in $M_2$ and yields an unsuitable recommendation.}
  \label{fig:intro}
  \Description{The user searches for hotel recommendations for a seaside getaway. Memory M1 identifies a solo trip. M2 updates the user's solo-vacation preference to quiet places while retaining a preference for lively places with friends. M3 records an older general preference for lively places. Panel I contrasts stopping at M1 with retrieving M2 to jointly support a quiet hotel. Panel II contrasts misapplying the older M3 with using the updated M2 for the current solo trip.}
\end{figure}

Figure~\ref{fig:intro} illustrates two difficulties in deriving applicable preferences from conversation histories. First, preference evidence is fragmented across conversations~\citep{maharana-etal-2024-evaluating}. A retrieved fragment often supplies partial support together with a clue for locating the missing information (Figure~\ref{fig:intro}, I). Second, interpreting these fragments requires their temporal and situational scope. An explicit update changes the applicability of an earlier preference~\citep{wu2025longmemeval}. A preference expressed for one situation does not necessarily hold for the current request~\citep{adomavicius2005context} (Figure~\ref{fig:intro}, II). These difficulties connect memory discovery with preference interpretation. Reading the available fragments reveals what remains unresolved, and further retrieval supplies evidence for a more complete assessment of the applicable preference~\citep{trivedi2023ircot,jiang2023flare}.

We propose \textbf{Weaver}, a personalized RAG framework built on a \textbf{Dual-Layer Retrieval and Reasoning Graph}. \textbf{Dependency Information Gap Tracking} maintains unresolved subquestions and their prerequisite dependencies in the upper graph. Each executable information need guides \textbf{Memory-Conditioned Incremental Retrieval}, which follows established premises and clues from retrieved conversations in the lower graph. \textbf{Scope-Aware Compositional Reasoning} interprets the accumulated fragments within their temporal and situational scopes. Complementary fragments form joint support, while independently sufficient evidence provides alternative support. The module also assesses conflicting claims within overlapping scopes. New evidence updates preference conclusions and the corresponding premises in the upper graph. Remaining information gaps guide subsequent retrieval.

\noindent\textbf{Our contributions are summarized as follows.}
\begin{itemize}
  \item \textbf{Incremental Memory Discovery.} We introduce a dual-layer graph that connects unresolved information needs to memory-conditioned retrieval trajectories. Established premises and source-text clues guide subsequent searches for missing evidence.
  \item \textbf{Scope-Aware Compositional Reasoning.} We represent joint evidential support together with its contextual applicability. Reassessing this support as evidence accumulates updates preference conclusions and the information needs they leave unresolved.
  \item \textbf{Experimental Evaluation.} Weaver achieves 86.42 ACC on PersonaMem~\citep{personamem2025} and 91.10 ACC on PrefEval~\citep{prefeval2025}, improving over the strongest evaluated baselines by 16.01 and 7.56 percentage points, respectively.
\end{itemize}

\section{Related Work}
\label{sec:related_work}

\noindent\textbf{\textit{Structured Memory for Personalization.}}
Personalized RAG retrieves user-specific records to condition generation. LaMP studies retrieval from personal profiles, including recency-aware selection~\citep{salemi2024lamp}. Subsequent work optimizes personal-document retrieval using feedback from downstream generation~\citep{salemi2024optimization}. Structured memory further represents connections within user histories. EMG-RAG organizes personal information in an editable memory graph~\citep{wang2024emg}, while MemWeaver combines behavioral and cognitive memories to capture temporal changes and semantic relationships~\citep{yu2026memweaver}. These representations support access to historical information, yet a stored relation does not necessarily provide all the evidence required for a specific request. Interpreting a retrieved fragment can depend on contextual information in another conversation. That interpretation exposes missing premises and search clues that were unavailable to the initial query.

\noindent\textbf{\textit{Iterative Retrieval for Personalization.}}
Building on stored memory representations, iterative retrieval uses intermediate results to determine subsequent memory access. RF-Mem selects between direct retrieval and iterative expansion in embedding space~\citep{zhang2026rfmem}. PGR forms retrieval probes through prospective reasoning and refines later probes with retrieved facts~\citep{chopra2026pgr}. PR$^2$ learns policies that interleave reasoning with retrieval from personal context~\citep{amirizaniani2026pr2}. Weaver examines how retrieved evidence establishes a preference applicable to the current request. It links unresolved information needs to the premises used by each retrieval step. Joint interpretation checks how fragments support a conclusion within its temporal and situational scope. The resulting conclusions update the information needs that guide subsequent searches.

\section{Method}
\label{sec:method}

\begin{figure*}[t]
  \centering
  \includegraphics[width=\textwidth]{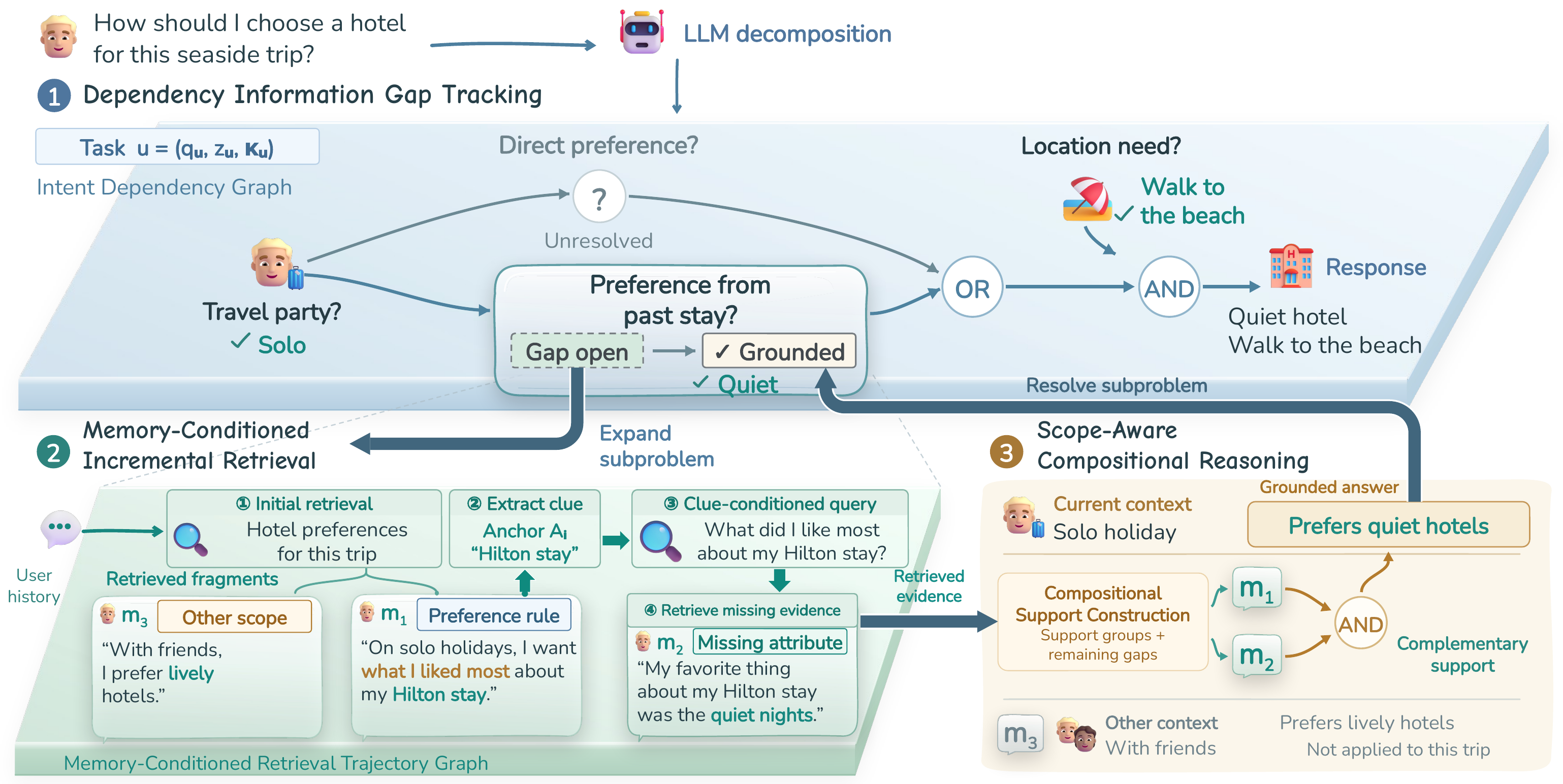}
  \caption{Overview of Weaver. (1) An unresolved subquestion in the upper
  Intent Dependency Graph expands into a lower retrieval trajectory.
  (2) Retrieved fragments expose clues that guide searches for missing
  evidence. (3) Compositional reasoning combines the fragments under the
  current context and returns a supported conclusion to resolve the
  originating subquestion and update the remaining gaps.}
  \label{fig:method}
  \Description{The upper graph decomposes the request into dependent subquestions. The selected subquestion expands downward into a retrieval trajectory, where retrieved fragments supply clues for further searches. Retrieved evidence flows to the reasoning module, which combines complementary fragments under the current context. The supported conclusion returns to the selected upper-graph node, changing its state from gap open to grounded.}
\end{figure*}

As illustrated in Figure~\ref{fig:method}, Weaver operates on a
Dual-Layer Retrieval and Reasoning Graph. The upper Intent Dependency
Graph represents task dependencies and tracks unresolved information
needs, while the lower Memory-Conditioned Retrieval Trajectory Graph
records how established premises and source-text clues guide incremental
memory discovery. Scope-Aware Compositional Reasoning
jointly interprets the retrieved evidence and updates task conclusions
and information gaps, guiding subsequent retrieval.

Algorithms~\ref{alg:gap-tracking}--\ref{alg:compositional-reasoning}
summarize this cycle. After initialization, available evidence is
interpreted before another query is issued. Gap tracking identifies an
executable need, retrieval adds evidence, and composition updates
conclusions and local gaps. These updates return to gap tracking,
allowing subquestion results to support the original request.
The response requirements are met only when the required outputs have
sufficient applicable support and no unresolved conflict; an empty
retrieval frontier alone does not establish completion.

\subsection{Dependency Information Gap Tracking}
\label{sec:gap-tracking}

In the upper graph, Weaver uses an LLM to decompose the current request
into dependent subquestions and tracks their unresolved information
needs as evidence accumulates.

\paragraph{Intent Dependency Graph Construction.}
Given a request $q$ and its known context $\kappa_q$, Weaver prompts
an LLM to identify the information needed for a personalized response
and decompose it into subquestions. The model determines how the answer
to one subquestion contributes to resolving another and organizes these
dependencies into an Intent Dependency Graph $G_I=(V_I,E_I)$. This
graph is a directed acyclic graph (DAG), with edges pointing from
prerequisite tasks to the tasks that use their outputs.

Each node $u=(q_u,z_u,\kappa_u)$ represents a subquestion $q_u$ with
an output variable $z_u$ to be resolved within scope $\kappa_u$. An
edge $(v,u)\in E_I$ indicates a potential premise dependency of
$u$ on the output of $v$. The LLM represents the dependencies of task $u$
through a logical expression $\varphi_u$ using conjunction and disjunction with negation.
As predecessor conclusions become available, this expression determines
which information remains necessary for downstream reasoning. We
assume that the initial graph covers the dependencies required by the
request. Subsequent reasoning updates task conclusions and their
relevance within this graph.

Let $\theta_t$ denote the established conclusions at iteration $t$,
initialized with information explicitly provided by the request and its
known context. A conclusion can serve as a premise when it has
sufficient support and no unresolved conflict in the required context.
Each step of evidence acquisition and interpretation for task $u$
uses context
$C_t^u=(q_u,\kappa_u,\gamma_t^u)$, where
$\gamma_t^u\subseteq\theta_t$ contains the established conclusions
on which that step relies. The graph records potential dependencies.
The step records its actual premises in $\gamma_t^u$. Different
information needs within a task can therefore rely on different
subsets of established conclusions.

\paragraph{Information Gap Tracking.}
Scope-Aware Compositional Reasoning returns a local gap set
$\delta_t^u$ describing what remains necessary to establish a
sufficiently supported conclusion for task $u$. Before evidence is retrieved,
these gaps are initialized from the unresolved task output and the
inputs needed to locate relevant information. Reading then makes the
gaps more specific.

Starting from the response requirements, Weaver follows the logical
dependencies in $G_I$ under the current conclusions to identify the
relevant task set $U_t$. This set contains tasks relevant to the current response
requirements and the upstream tasks needed to supply their inputs.
The outstanding information needs are
\begin{equation}
\Delta_t
=\left\{(u,d)\mid u\in U_t,\ d\in\delta_t^u\right\}.
\label{eq:information-gaps}
\end{equation}
An information need remains in $\Delta_t$ when a required input is
unavailable. Weaver first advances the upstream task responsible for
that input, allowing the dependent retrieval step to proceed once the
premise is established.

Let $R_t^{u,d}$ denote the prerequisite input slots for investigating
gap $d$, determined by $\varphi_u$ and the current conclusions. Let
$A_t^u$ collect input slots with values in $\theta_t$ applicable to
task $u$, and let $b_t^{u,d}=1$ indicate an available, unexecuted
retrieval query, including a clue-based probe. The executable frontier is
\begin{equation}
\mathcal{F}_t
=\left\{(u,d)\in\Delta_t\mid
R_t^{u,d}\subseteq A_t^u,\ b_t^{u,d}=1\right\}.
\label{eq:retrieval-frontier}
\end{equation}
An initial query is formed from a gap $(u,d)\in\mathcal F_t$ and
its established premises. Later queries can also incorporate
source-text clues from retrieved fragments.

After conclusion revision, Algorithm~\ref{alg:gap-tracking} refreshes
the relevant tasks and retrieval frontier. Supported conclusions enable
downstream retrieval; lost support or unresolved conflicts reopen gaps.
Unnecessary tasks are paused with their evidence and dependencies
retained, so later findings can reactivate them.

\begin{weaveralgorithm}{Dependency Information Gap Tracking}{alg:gap-tracking}
\alginput{Request $q$, context $\kappa_q$, and current graph, conclusion, gap, and query states.}
\algoutput{Relevant tasks $U_t$, gaps $\Delta_t$, and frontier $\mathcal F_t$.}
\begin{weaversteps}
  \item \textbf{if} $G_I$ is uninitialized \textbf{then}
  \item \hspace*{0.7em} Decompose $q$ into subquestions with an LLM; build their prerequisite DAG $G_I$ and initialize $\theta_t$ and local gaps.
  \item \textbf{end if}
  \item Trace response dependencies under $\theta_t$ to obtain $U_t$, including upstream tasks supplying missing inputs.
  \item Collect relevant local gaps into $\Delta_t$ (Eq.~\ref{eq:information-gaps}).
  \item Form $\mathcal F_t$ from gaps with available inputs and unexecuted queries (Eq.~\ref{eq:retrieval-frontier}).
  \item Pause tasks outside $U_t$; retain their evidence and dependencies for possible reactivation.
  \item \textbf{return} $U_t,\Delta_t,\mathcal F_t$.
\end{weaversteps}
\end{weaveralgorithm}

\subsection{Memory-Conditioned Incremental Retrieval}
\label{sec:incremental-retrieval}

Weaver incrementally retrieves missing evidence for the information
needs identified by the upper graph and records the discovery process
in the lower graph. As conversations are retrieved and interpreted,
newly exposed clues guide further searches through the user's
interaction history. Memory conditioning refers to the use of retrieved
source-text clues and memory-derived premises when constructing later
queries. The current request and its scope remain part of the query context.

\paragraph{Clue-Conditioned Retrieval.}
At reasoning state $t$, the first module provides $\mathcal F_t$.
A selected pair $(u,d)\in\mathcal F_t$ identifies the subquestion
and information gap to investigate. For retrieval event $i$, the
context $C_t^u=(q_u,\kappa_u,\gamma_t^u)$ supplies the subquestion,
its applicability conditions, and the established premises on which
this search relies.

Previously retrieved fragments can expose clues for locating missing
information. Weaver retains each selected source-text span and its
source reference as an anchor. Let $\mathcal A_i$ contain the anchors
actually used to formulate query $p_i$. The LLM query-generation
process $\mathcal Q$ combines the current context, selected gap, and
these anchors:
\begin{equation}
p_i=\mathcal Q\!\left(C_t^u,d,\mathcal A_i\right).
\label{eq:conditioned-query}
\end{equation}
An initial query without source-text clues has $\mathcal A_i=\varnothing$.
For a clue-conditioned query, the LLM retains the locating terms from
the anchors and asks for the information needed to resolve $d$ under
the premises in $\gamma_t^u$. Each anchor preserves its original
wording and source reference, linking the query to the fragment that
prompted it.

The search space $H_{a,q}$ contains the conversation fragments of user
$a$ accessible when request $q$ is made. The existing retriever maps
queries and fragments into representations through encoders $h_Q$
and $h_M$. Its similarity function $\operatorname{sim}$ scores each
fragment $m\in H_{a,q}$. The operator $\operatorname{TopK}$ returns
the fragments with the highest scores, up to a maximum of $K$:
\begin{equation}
\mathcal R_i
=
\operatorname{TopK}_{m\in H_{a,q}}
\operatorname{sim}\!\left(h_Q(p_i),h_M(m)\right).
\label{eq:memory-conditioned-retrieval}
\end{equation}
Initial and clue-conditioned queries share this retrieval backend
without an additional reranker. Their returned fragments accumulate
as evidence for the selected task. Each executed query is recorded
together with its context and premises, providing the query state
needed to update $b_t^{u,d}$ in Eq.~\ref{eq:retrieval-frontier}.

\paragraph{Retrieval Trajectory Construction.}
Weaver records the discovery process in the Memory-Conditioned
Retrieval Trajectory Graph $G_t^R=(V_t^R,E_t^R)$. Here, $V_t^R$
contains the retrieval steps recorded by iteration $t$, and $E_t^R$
contains their directed discovery relationships. Each node
$x_i\in V_t^R$ represents an executed retrieval step. It records which
task and gap motivated the query and retains the context under which
the query was formed. The query $p_i$ and its anchors $\mathcal A_i$
are linked to the returned fragments $\mathcal R_i$. The recorded
task associates the node with the upper graph.

Let $\operatorname{src}(\mathcal A_i)$ collect the earlier retrieval
nodes from which the anchors in query $p_i$ are taken. For recorded
nodes $x_j,x_i\in V_t^R$, the discovery relationship is
\begin{equation}
(x_j,x_i)\in E_t^R
\iff j<i\ \land\ x_j\in\operatorname{src}(\mathcal A_i).
\label{eq:retrieval-edges}
\end{equation}
An edge therefore records an actual contribution of source-text clues
to a later query. If several retrieval steps supply anchors for the
same query, all corresponding edges are retained. The temporal order
$j<i$ makes the trajectory graph acyclic.

Algorithm~\ref{alg:incremental-retrieval} records this process. Fragments
can be shared across tasks, while queries under different premises
retain distinct records. A fragment from another situation may provide
a locating clue without supporting the current preference; composition
assesses its applicability and reinterprets accumulated evidence when
premises change.

\begin{weaveralgorithm}{Memory-Conditioned Incremental Retrieval}{alg:incremental-retrieval}
\alginput{Gap $(u,d)\in\mathcal F_t$, context $C_t^u$, history $H_{a,q}$, evidence $\mathcal E_t^u$, trajectory $G_t^R$, and query records.}
\algoutput{Fragments $\mathcal R_i$; updated evidence, trajectory, and query records.}
\begin{weaversteps}
  \item Gather required premises in $C_t^u$; select source-text anchors $\mathcal A_i$ for missing support, or $\varnothing$ if none are available.
  \item Form $p_i=\mathcal Q(C_t^u,d,\mathcal A_i)$.
  \item Retrieve $\mathcal R_i$ from $H_{a,q}$ (Eq.~\ref{eq:memory-conditioned-retrieval}); accumulate $\mathcal E^u\leftarrow\mathcal E_t^u\cup\mathcal R_i$.
  \item Record node $x_i$ with its task, gap, context, anchors, query, and returned fragments.
  \item Link each source node $x_j\in\operatorname{src}(\mathcal A_i)$ to $x_i$ in the trajectory.
  \item Mark $p_i$ as executed under its recorded context and premises.
  \item \textbf{return} $\mathcal R_i$, updated evidence, trajectory, and query records.
\end{weaversteps}
\end{weaveralgorithm}

\subsection{Scope-Aware Compositional Reasoning}
\label{sec:evidence-synthesis}

As evidence accumulates, Weaver jointly interprets the retrieved fragments
within their temporal and situational scopes. It establishes conclusions
with sufficient applicable support and revises them when that support
changes. The resulting conclusions and
unresolved information needs are returned to the upper graph to guide
subsequent retrieval.

\paragraph{Compositional Support Construction.}
Let $\mathcal E_t^u$ contain the conversation fragments accumulated for
task $u$. Weaver prompts an LLM to interpret these fragments under
$C_t^u$, resolving references across conversations and identifying
which memories and established premises jointly support each
candidate conclusion. The output records the conclusion, its
supporting sources, and its applicability conditions. Unresolved
references and unclear applicability conditions enter the local gap
set $\delta_t^u$.

A candidate conclusion is $f=(z_u,v,\kappa_f)$, where $v$ is a proposed
value of the task output and $\kappa_f$ specifies its applicability.
A support group is recorded as $s=(f,M_s,\Gamma_s,\kappa_s)$.
The fragments $M_s\subseteq\mathcal E_t^u$ and premises
$\Gamma_s\subseteq\gamma_t^u$ jointly support $f$ under the conditions
in $\kappa_s$. Members of a group can supply complementary information,
so an individual memory need not support $f$ on its own. Removing a
required member leaves that group's justification incomplete.

For each group, the LLM assesses whether the combined interpretation
is sufficient for $f$, whether its temporal and situational conditions
apply to the task, and whether its required premises remain
established. The validity indicator $V_t(s,u)$ records this judgment,
with value~1 when these requirements hold and~0 otherwise. In
particular, shared topic alone does not establish applicability, and
an unresolved reference leaves the support incomplete. Let $S_t(f)$
collect the support groups for $f$. Distinct complete groups provide
alternative justifications:
\refstepcounter{equation}\label{eq:alternative-support}%
$\operatorname{Sup}_t(f,u)=\bigvee_{s\in S_t(f)}V_t(s,u)$~(\theequation).
This relation aggregates the LLM's judgments of whole support groups.
A conclusion retains support while at least one group remains valid.

\paragraph{Contextual Conclusion Revision.}
Weaver compares supported conclusions for the same task output within
overlapping applicability conditions. An unresolved conflict occurs
when the LLM judges their values incompatible within that shared
scope. Semantically equivalent values are merged. Different values
can coexist under disjoint conditions. When a subquestion asks about
multiple situations, its answer can contain several compatible
scoped conclusions. Completeness is assessed against the requested
output, and the task remains open while required information is
missing or a relevant conflict is unresolved.

The LLM reassesses conclusions when evidence provides an explicit
correction or clarifies applicability. A later timestamp alone does
not establish a preference change. Changes to evidence and premises
also trigger reassessment of the support groups that depend on them.
An alternative valid group can preserve a conclusion when another
group loses support.

Let $\mathcal W_t\subseteq\theta_t$ contain established conclusions
that lose sufficient applicable support or acquire an unresolved
conflict. Let $\mathcal N_t$ contain newly established or revised
scoped conclusions with sufficient support and no unresolved conflict.
Reassessment follows affected premise dependencies in dependency order,
including downstream conclusions that rely on a changed premise.
Weaver completes this reassessment before finalizing $\mathcal N_t$
and $\mathcal W_t$ and returning to retrieval. The updated conclusion
set is
\begin{equation}
\theta_{t+1}
=\left(\theta_t\setminus\mathcal W_t\right)\cup\mathcal N_t.
\label{eq:conclusion-update}
\end{equation}
Algorithm~\ref{alg:compositional-reasoning} returns revised conclusions
and local gaps to the upper graph, which refreshes premises and
recomputes $U_{t+1}$, $\Delta_{t+1}$, and $\mathcal F_{t+1}$.
Retrieval edges record discovery: changing a locating clue does not
invalidate conclusions supported independently of it.

\begin{weaveralgorithm}{Scope-Aware Compositional Reasoning}{alg:compositional-reasoning}
\alginput{Context $C_t^u$, evidence $\mathcal E_t^u$, support groups, and established conclusions $\theta_t$.}
\algoutput{Updated support groups, $\theta_{t+1}$, and affected local gaps.}
\begin{weaversteps}
  \item Interpret $\mathcal E_t^u$ under $C_t^u$; resolve references, preference subjects, and applicability conditions.
  \item Construct or revise support groups $s=(f,M_s,\Gamma_s,\kappa_s)$ for candidate conclusions.
  \item Assess group validity $V_t(s,u)$ and aggregate alternative support for each conclusion (Eq.~\ref{eq:alternative-support}).
  \item Merge equivalent conclusions; identify incompatible values within overlapping scopes.
  \item Propagate proposed premise changes in dependency order, reassessing affected support groups and downstream conclusions.
  \item After reassessment, collect conclusions to establish or revise in $\mathcal N_t$, and those losing support or acquiring unresolved conflicts in $\mathcal W_t$.
  \item Update $\theta_{t+1}=(\theta_t\setminus\mathcal W_t)\cup\mathcal N_t$.
  \item Update affected local gaps for missing support, unresolved references, unclear applicability, or conflicts.
  \item \textbf{return} updated support groups, $\theta_{t+1}$, and local gaps to gap tracking.
\end{weaversteps}
\end{weaveralgorithm}

\section{Experiments}
\label{sec:experiments}
\setlength{\textfloatsep}{8pt plus 2pt minus 2pt}
\setlength{\dbltextfloatsep}{8pt plus 2pt minus 2pt}
\setlength{\intextsep}{6pt plus 2pt minus 2pt}

We evaluate whether Weaver derives applicable preferences from fragmented user histories. RQ1 compares personalized response quality with existing memory-augmented methods. RQ2 measures the contribution of each module. RQ3 examines the computational cost of the observed personalization gains. RQ4 evaluates the discovery and composition of preference-relevant memories within user histories, followed by an evaluation beyond personalization.

\subsection{Experimental Setup}
\label{sec:experimental-setup}

\paragraph{Datasets.}
PersonaMem~\citep{personamem2025} evaluates personalized responses using evolving user information; we use its 32K-history configuration. PrefEval~\citep{prefeval2025} evaluates preference following in conversational contexts. Section~\ref{sec:evidence-reasoning} additionally evaluates MuSiQue~\citep{trivedi-etal-2022-musique} and WebWalkerQA~\citep{wu-etal-2025-webwalker} beyond user histories.

\begin{table}[!t]
\centering
\fontsize{7.2}{8.8}\selectfont
\setlength{\tabcolsep}{1.4pt}
\caption{Main results on personalized response generation. All scores are percentages. Higher is better. Best results are bold and second-best results are underlined.}
\label{tab:main-results}
\begin{tabular*}{\columnwidth}{@{\extracolsep{\fill}}llrrrrrr@{}}
\toprule
& & \multicolumn{3}{c}{PersonaMem~\citep{personamem2025}} & \multicolumn{3}{c}{PrefEval~\citep{prefeval2025}} \\
\cmidrule(lr){3-5}\cmidrule(lr){6-8}
Category & Method & ACC & F1 & EM & ACC & F1 & EM \\
\midrule
Ours & Weaver & \textbf{86.42} & \textbf{78.57} & \textbf{73.84} & \textbf{91.10} & \textbf{85.37} & \textbf{82.61} \\
\midrule
\multirow{4}{*}{\shortstack[l]{Memory\\Organization}} & MemCoE~\citep{xu-etal-2026-learning} & \underline{70.41} & \underline{68.06} & \underline{63.31} & 82.40 & 81.30 & 78.14 \\
& BMAM~\citep{li-etal-2026-bmam} & 61.67 & 59.90 & 54.12 & 74.17 & 72.90 & 69.83 \\
& MemoryBank~\citep{zhong2024memorybank} & 62.48 & 60.72 & 55.05 & \underline{83.54} & \underline{82.16} & \underline{79.08} \\
& EMG-RAG~\citep{wang2024emg} & 57.92 & 55.83 & 49.91 & 78.11 & 74.85 & 71.62 \\
\midrule
\multirow{2}{*}{\shortstack[l]{Preference\\RAG}} & RF-Mem~\citep{zhang2026rfmem} & 64.21 & 63.50 & 58.76 & 79.62 & 76.11 & 73.25 \\
& PGR~\citep{chopra2026pgr} & 60.12 & 58.19 & 53.68 & 77.84 & 74.33 & 71.49 \\
\bottomrule
\end{tabular*}
\end{table}

\paragraph{Baselines.}
We group six baselines by their primary focus. Memory organization includes MemCoE~\citep{xu-etal-2026-learning}, which learns organization and update policies. BMAM~\citep{li-etal-2026-bmam} divides memory into specialized subsystems. MemoryBank~\citep{zhong2024memorybank} maintains conversational memory through forgetting and reinforcement. EMG-RAG~\citep{wang2024emg} stores personal information in an editable graph. Preference RAG includes RF-Mem~\citep{zhang2026rfmem}, whose recollection path progressively expands evidence. PGR~\citep{chopra2026pgr} refines prospective retrieval probes with retrieved facts. RF-Mem and PGR also provide the reported comparisons on MuSiQue and WebWalker.

\paragraph{Evaluation Metrics.}
On PersonaMem and PrefEval, ACC measures option-label correctness; EM and token F1 are auxiliary text-match scores comparing the generated response with the full gold-option text. Appendix~\ref{sec:response-evaluation} separates these outputs. MuSiQue uses F1, EM, and supporting-evidence recall within the top five results (R@5); WebWalker uses ACC only. Scores are percentages; differences are absolute percentage points unless labeled as relative changes.

\paragraph{Implementation.}
Weaver and all six baselines use Qwen3.8-27B, Qwen3-Embedding-8B, and the same 16,384-token context limit. Appendix~\ref{sec:implementation-details} provides hardware, decoding, and evaluation details.

\subsection{Main Results (RQ1)}
\label{sec:main-results}

Table~\ref{tab:main-results} reports personalized response quality across both baseline groups. Weaver achieves the highest ACC on both benchmarks, with corresponding gains in F1 and EM.

On PersonaMem, MemCoE is the strongest baseline across all three metrics. Weaver improves ACC from 70.41 to 86.42, F1 from 68.06 to 78.57, and EM from 63.31 to 73.84, gains of 16.01, 10.51, and 10.53 points. On PrefEval, MemoryBank is the strongest baseline across all three metrics. Weaver improves ACC from 83.54 to 91.10 and F1 from 82.16 to 85.37. EM increases from 79.08 to 82.61. These gains amount to 7.56, 3.21, and 3.53 percentage points, respectively. The gains are larger on PersonaMem. Section~\ref{sec:evidence-reasoning} examines whether Weaver finds sufficient support and uses it effectively.

\begin{figure}[!t]
\centering
\includegraphics[width=\columnwidth]{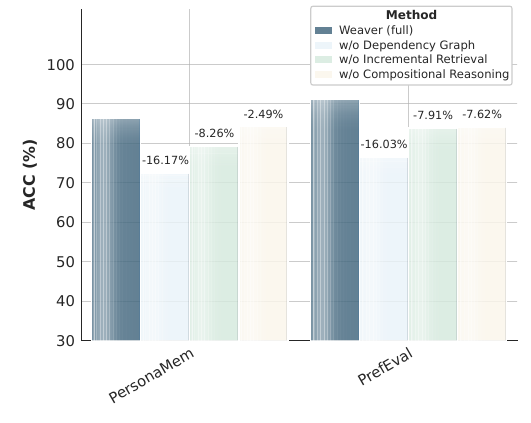}
\caption{ACC ablations on PersonaMem and PrefEval. Dark blue shows Weaver, and the lighter bars show independent module removals. Labels indicate relative ACC decreases from Weaver in percent.}
\label{fig:ablation}
\Description{A grouped bar chart for PersonaMem and PrefEval. Each group contains the complete Weaver model and variants without the dependency graph, incremental retrieval, and compositional reasoning. Dark blue denotes the full model. Pale blue, green, and cream denote removal of the dependency graph, incremental retrieval, and compositional reasoning, respectively. Labels show relative percentage decreases from the full-model ACC score.}
\end{figure}

\subsection{Ablation Study (RQ2)}
\label{sec:ablation}

We evaluate three independent ablations, each removing one module from the complete framework. Figure~\ref{fig:ablation} shows ACC on the two personalization benchmarks, and Appendix~\ref{sec:complete-ablation} reports all metrics on all four datasets.
\begin{itemize}
  \item \textit{w/o Dependency Graph} removes the upper graph. The original question is passed directly to modules~2 and~3, with subquestion decomposition and upper-graph gap tracking disabled.
  \item \textit{w/o Incremental Retrieval} replaces Memory-Conditioned Incremental Retrieval with direct dense retrieval for each subquestion. It retains the upper-graph subquestions and uses module~3 to combine the retrieved evidence.
  \item \textit{w/o Compositional Reasoning} removes Scope-Aware Compositional Reasoning. It retains the dependency graph and incremental retrieval, and passes all evidence accumulated in the lower graph directly to the reader.
\end{itemize}

The complete framework outperforms every ablation on all reported metrics. Removing the dependency graph produces the largest ACC decrease on the personalization benchmarks and WebWalker, and the largest F1 decrease on MuSiQue.

\paragraph{Dependency graph.}
Applying modules~2 and~3 directly to the original question reduces ACC by 13.97 points on PersonaMem and 14.60 points on PrefEval. MuSiQue F1 decreases by 11.20 points, and WebWalker ACC by 8.71 points. These results support organizing memory discovery around dependent subquestions and their unresolved information needs. This ablation removes decomposition together with dependency-based gap tracking, measuring their combined contribution.

\begin{figure}[t]
    \centering
    \includegraphics[width=\columnwidth]{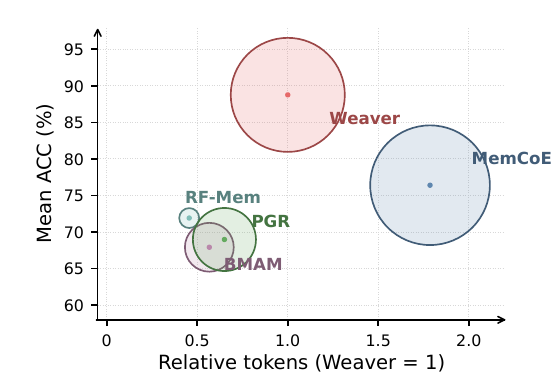}
    \caption{Personalization quality and computational cost for the five methods with available cost measurements. Points show mean ACC over PersonaMem and PrefEval against relative token consumption, with Weaver normalized to 1. Circle area encodes mean runtime. Higher positions indicate better accuracy, and positions farther left indicate lower token cost.}
    \label{fig:efficiency}
    \Description{A bubble plot comparing RF-Mem, BMAM, PGR, Weaver, and MemCoE. Horizontal position is token consumption relative to Weaver, vertical position is mean personalization accuracy, and circle area represents mean runtime. The relative token values are approximately 0.46 for RF-Mem, 0.57 for BMAM, 0.65 for PGR, 1.00 for Weaver, and 1.79 for MemCoE. Weaver has the highest mean accuracy.}
\end{figure}

\paragraph{Incremental retrieval.}
Replacing module~2 with direct dense retrieval reduces ACC by 7.14 points on PersonaMem and 7.21 points on PrefEval. MuSiQue F1 decreases by 4.86 points, and WebWalker ACC by 1.49 points. Since this variant retains compositional reasoning, the comparison shows that combining directly retrieved evidence does not recover the performance of incremental memory discovery guided by retrieved clues.

\paragraph{Compositional reasoning.}
Passing all lower-graph evidence directly to the reader lowers ACC from 86.42 to 84.27 on PersonaMem and from 91.10 to 84.16 on PrefEval, decreases of 2.15 and 6.94 points. MuSiQue F1 decreases by 4.31 points, and WebWalker ACC by 0.33 points. Because module~3 also updates conclusions and remaining gaps during retrieval, this ablation measures its contribution to the iterative process as well as to evidence interpretation.

\subsection{Personalization Quality and Cost (RQ3)}
\label{sec:efficiency}

Figure~\ref{fig:efficiency} compares the five methods with available cost measurements using mean ACC over PersonaMem and PrefEval, with approximate aggregate token ratios normalized to Weaver.
The relative token costs are approximately 0.46 for RF-Mem, 0.57 for BMAM, 0.65 for PGR, 1.00 for Weaver, and 1.79 for MemCoE.
Weaver achieves the highest mean ACC at 88.76, compared with 71.92 for RF-Mem, 67.92 for BMAM, 68.98 for PGR, and 76.41 for MemCoE.
The three lower-cost baselines use approximately 46--65\% of Weaver's tokens, illustrating a trade-off between token expenditure and personalization quality.
Compared with MemCoE, Weaver uses approximately 56\% as many tokens, a reduction of about 44\%, while improving mean ACC by 12.36 percentage points, rounded after subtraction.
Mean end-to-end latency per request is 327\,s for Weaver and 360\,s for MemCoE.
On PrefEval, Weaver also exceeds MemCoE by 8.70 ACC points and 4.07 F1 points.

Weaver reinterprets available evidence before further retrieval, directing computation toward the support needed for the current request.

\subsection{Preference Memory Discovery and Composition (RQ4)}
\label{sec:evidence-reasoning}

How effectively does Weaver discover and compose preference-relevant memories to answer the current request? Such memories include explicit preferences, past experiences, evaluations, and contextual information whose relevance depends on the request. We first evaluate whether retrieval recovers sufficient support, then assess composition given identical candidate memories, and finally examine applicability beyond personalization.

\begin{table}[t]
\centering
\caption{Memory discovery and composition beyond personalization. All scores are percentages. Best results are bold and second-best results are underlined.}
\label{tab:reasoning-transfer}
\fontsize{8}{9.5}\selectfont
\setlength{\tabcolsep}{3pt}
\begin{tabular*}{\columnwidth}{@{\extracolsep{\fill}}lrrrc@{}}
\toprule
& \multicolumn{3}{c}{MuSiQue~\citep{trivedi-etal-2022-musique}} & WebWalker~\citep{wu-etal-2025-webwalker} \\
\cmidrule(lr){2-4}\cmidrule(lr){5-5}
Method & F1 & EM & R@5 & ACC \\
\midrule
Weaver & \textbf{69.52} & \textbf{59.36} & \textbf{85.22} & \textbf{60.31} \\
RF-Mem~\citep{zhang2026rfmem} & 65.81 & 55.92 & 82.72 & 56.22 \\
PGR~\citep{chopra2026pgr} & \underline{67.11} & \underline{57.43} & \underline{84.31} & \underline{59.02} \\
\bottomrule
\end{tabular*}
\end{table}

\begin{table}[!t]
\centering
\caption{Complete support recovery (\%), averaged over three runs. Groups indicate the minimum memories needed for sufficient support. IR: Incremental Retrieval. $\Delta$: Weaver's gain (percentage points), rounded after subtraction.}
\label{tab:evidence-recovery}
\fontsize{8.5}{10}\selectfont
\setlength{\tabcolsep}{3pt}
\renewcommand{\arraystretch}{1.18}
\begin{tabularx}{\columnwidth}{>{\raggedright\arraybackslash}Xrrr}
\toprule
\textbf{Support requirement} & \textbf{w/o IR} & \textbf{Weaver} & \textbf{$\Delta$ (pp)} \\
\midrule
\rowcolor{RQ4RowOne}
One memory suffices & 94.17 & \textbf{95.83} & +1.67 \\
\rowcolor{RQ4RowTwo}
Two memories required & 81.67 & \textbf{87.50} & +5.83 \\
\rowcolor{RQ4RowThree}
$\geq$3 memories required & 72.50 & \textbf{81.67} & +9.17 \\
\bottomrule
\end{tabularx}
\end{table}

\paragraph{Preference memory discovery.}
Can Weaver recover sufficient support from preference-relevant memories distributed across user histories? We compare Weaver with \textit{w/o Incremental Retrieval}, which directly retrieves dense matches for each subquestion while retaining the dependency graph and compositional reasoning. We fix the backbone, retrieval backend, history segmentation, and cumulative retrieval budget. Recovery succeeds when memories accumulated across retrieval steps contain at least one complete sufficient evidence set for the current request; partial support counts as a miss. The two diagnostics use 120 questions from PersonaMem, grouped by the minimum number of memories needed for sufficient support, with 40 questions per group.

Table~\ref{tab:evidence-recovery} shows that direct dense retrieval already performs well when one memory suffices, recovering complete support for 94.17\% of questions compared with Weaver's 95.83\%. The gain widens to 5.83 points when two memories are required and 9.17 points when at least three are required. Clues found in earlier fragments guide subsequent searches for missing support. These results support memory-conditioned retrieval for recovering dispersed memories that jointly support a preference judgment. Improved recovery is accompanied by higher overall answer ACC across all 120 questions, from 79.17 to 84.72.

\begin{table}[!t]
\centering
\caption{Answer accuracy (\%) with identical candidate memories containing sufficient support and distractors, averaged over three runs. Groups indicate required support, not input size. $\Delta$: Weaver's gain (percentage points).}
\label{tab:compositional-reasoning}
\fontsize{8.5}{10}\selectfont
\setlength{\tabcolsep}{3pt}
\renewcommand{\arraystretch}{1.18}
\begin{tabularx}{\columnwidth}{>{\raggedright\arraybackslash}Xrrr}
\toprule
\textbf{Support requirement} & \textbf{Direct reader} & \textbf{Weaver} & \textbf{$\Delta$ (pp)} \\
\midrule
\rowcolor{RQ4RowOne}
One memory suffices & 89.17 & \textbf{91.67} & +2.50 \\
\rowcolor{RQ4RowTwo}
Two memories required & 82.50 & \textbf{87.50} & +5.00 \\
\rowcolor{RQ4RowThree}
$\geq$3 memories required & 72.50 & \textbf{80.83} & +8.33 \\
\bottomrule
\end{tabularx}
\end{table}

\paragraph{Preference memory composition.}
Does compositional reasoning improve personalized answer accuracy given the same candidate memories? We bypass retrieval and provide Weaver and the direct reader with identical candidate memories containing all required evidence and distractors. Supporting fragments are not identified, so the models must determine which records apply and how they support the current request. Direct reader answers from this candidate set; Weaver applies Scope-Aware Compositional Reasoning to interpret and combine the memories before answering. We fix the backbone, task premises, and memory order, with no additional retrieval. We compare answer accuracy within the same support-requirement groups.

Table~\ref{tab:compositional-reasoning} shows an ACC gain of 2.50 points when one memory suffices, increasing to 5.00 points when two are required and 8.33 points when at least three are required. In the last group, accuracy rises from 72.50 to 80.83 despite both methods having access to sufficient support. The widening advantage supports interpreting relationships among memories when several records must jointly support the answer. Same-topic preferences from other contexts remain in the input, requiring the models to consider applicability to the current request.

Together, these diagnostics show benefits in finding sufficient preference-relevant memories and combining their support for personalized answers. They do not separately quantify temporal-update handling or situational applicability.

\paragraph{Beyond preference memories.}
Does the same discovery and composition process extend beyond preference memories? We evaluate Weaver on MuSiQue~\citep{trivedi-etal-2022-musique} and WebWalkerQA~\citep{wu-etal-2025-webwalker}. MuSiQue tests multi-hop question answering through composed questions. WebWalkerQA, abbreviated as WebWalker, tests information seeking with Web pages accessed at runtime. Both tasks require locating and combining factual support outside user histories.

Table~\ref{tab:reasoning-transfer} shows improvements over PGR, the strongest reported baseline on both tasks. On MuSiQue, F1 rises from 67.11 to 69.52, EM from 57.43 to 59.36, and R@5 from 84.31 to 85.22. On WebWalker, ACC rises from 59.02 to 60.31. These gains support the applicability of the discovery and composition process beyond preference-related histories. As end-to-end results, they do not isolate the contributions of memory discovery and composition.

\begin{figure}[t]
    \centering
    \includegraphics[width=\columnwidth]{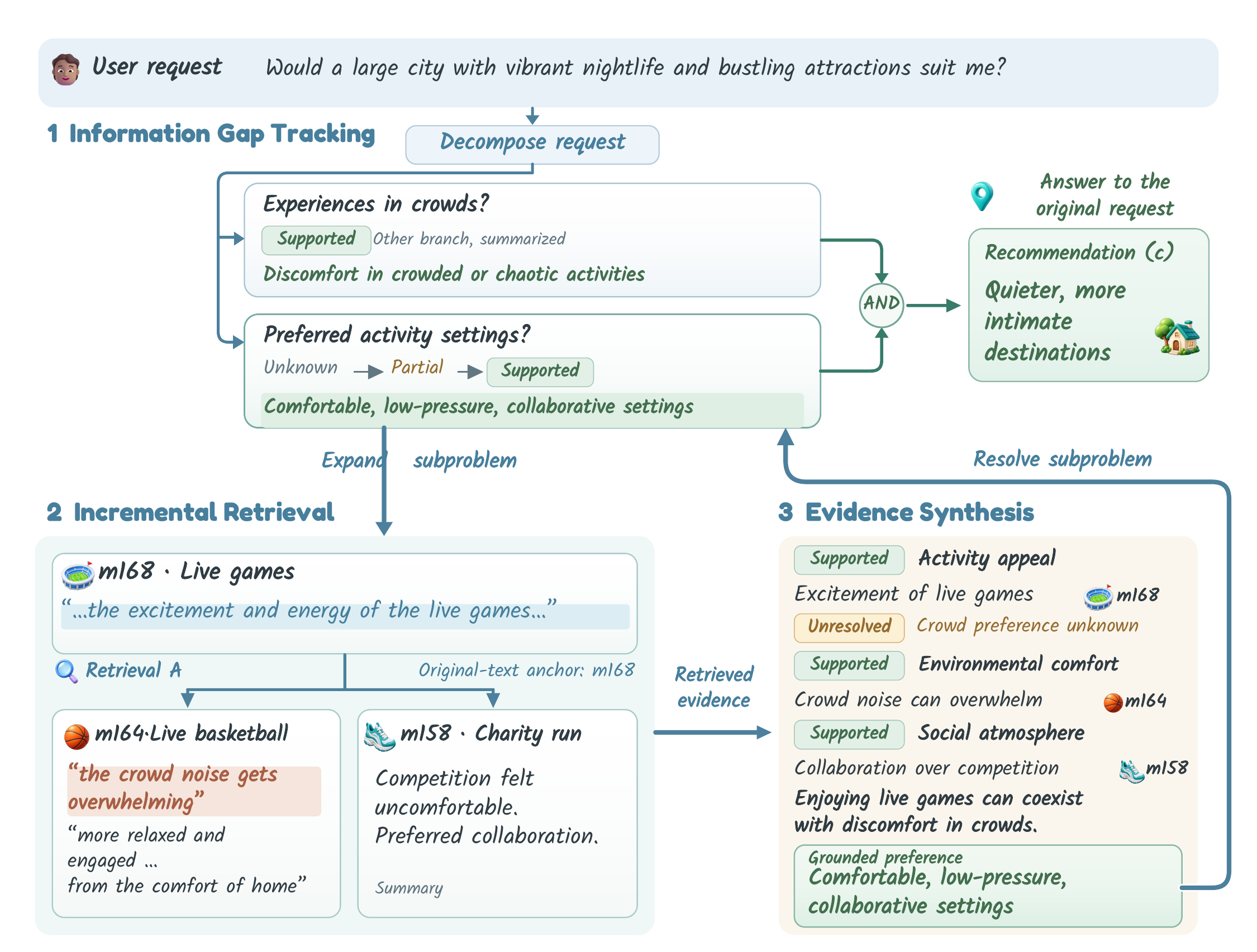}
    \caption{Case study on PersonaMem. The request is decomposed into activity-setting and crowd-experience subquestions. Selected retrieval steps recover memories that support a contextualized preference. This preference returns to resolve the activity-setting subquestion and combines with the other branch to answer the original request.}
    \label{fig:case-study}
    \Description{A historical Weaver development trace organized into information gap tracking, incremental retrieval, and evidence synthesis. The original request branches into preferred activity settings and experiences in crowds. The selected preference branch expands into retrieval using m168 as an original-text anchor. Memories m164 and m158 supply contextual evidence about crowd noise, home comfort, and collaboration. Evidence synthesis returns a grounded preference to the selected subquestion, which combines with the summarized crowd-experience branch through AND to recommend quieter, more intimate destinations. The figure selects steps from a 20-retrieval run and omits 16 earlier conditional probes and the secondary dance-class search. Supported and Unresolved labels are editorial evidence summaries.}
\end{figure}

\subsection{Case Study}
\label{sec:case-study}

We examine a PersonaMem example in which the user asks whether a large city with vibrant nightlife and bustling attractions would suit them. Figure~\ref{fig:case-study} shows selected steps from a historical development run of Weaver (v3.1). The upper graph separates the user's preferred activity settings from their experiences in crowded environments, and the recommendation depends on both. The preference task initially remains unresolved.

\paragraph{Incremental memory discovery.}
During later retrieval, a memory recalling the excitement and energy of live games (m168) provides an original-text anchor for further search. This search retrieves four additional documents, including a basketball experience (m164) describing overwhelming crowd noise and greater comfort watching at home, and a charity-run experience (m158) expressing discomfort with competition and a preference for collaboration. These newly retrieved experiences enter the subsequent evidence interpretation. Another new memory (m166 and m167), together with m168 as a premise, guides the next search to a dance-class experience (m156 and m157). All five additional documents appear in the final citations. The lower graph therefore records how previously retrieved text leads to further evidence while the upper preference task remains open.

\paragraph{Scope-aware compositional reasoning.}
The accumulated evidence distinguishes the appeal of an activity from the conditions under which the user enjoys it. Recalling the excitement of live games does not establish a preference for crowded environments. The basketball account expresses enjoyment alongside discomfort with noise. The charity-run account adds a social dimension, favoring collaboration over competition. Interpreting these experiences together allows enthusiasm for an activity to coexist with a preference for comfortable, low-pressure settings. After the accumulated evidence is read, including the dance-class experience, the preference task moves from partial to supported. The resulting premise is combined with the separately established summary of discomfort in crowded settings and chaotic activities. The final recommendation selects the correct option~(c), which favors quieter, more intimate destinations.

This recommendation transfers preferences from related activities to travel, since the available history contains no direct preference about cities and nightlife. The trace illustrates how evidence discovered in the lower graph is interpreted under its original context and returned as a premise for the upper graph's recommendation.


\section{Conclusion}
\label{sec:conclusion}

We presented Weaver, a personalized RAG framework that discovers and composes preference-relevant memories. Its dual-layer graph tracks dependent information needs and guides incremental retrieval using premises and clues from user histories. Compositional reasoning interprets evidence within applicable contexts, while updated conclusions and remaining gaps direct subsequent retrieval. Experiments on PersonaMem and PrefEval show improved answer quality, and controlled diagnostics show more complete support recovery and higher accuracy with identical candidate memories, especially when multiple memories are required. These results support coupling memory discovery with contextual preference reasoning for personalized responses.

\bibliographystyle{ACM-Reference-Format-CitationOrder}
\bibliography{references}

\appendix
\raggedbottom
\section{Complete Ablation Results}
\label{sec:complete-ablation}

Table~\ref{tab:ablation} reports the complete results for the independent ablations in Section~\ref{sec:ablation}. Each variant removes one module while retaining the other two.

\begingroup
\captionsetup{type=table}
\caption{Complete ablation results. All scores are percentages. Higher is better. Best results are bold and second-best results are underlined.}
\label{tab:ablation}
\par\nobreak
\endgroup

\noindent\begin{minipage}{\columnwidth}
\fontsize{8.5}{10.2}\selectfont
\setlength{\tabcolsep}{3pt}
\begin{tabular*}{\columnwidth}{@{\extracolsep{\fill}}lrrr@{}}
\toprule
\multicolumn{4}{c}{\textbf{PersonaMem}~\citep{personamem2025}} \\
Variant & ACC & F1 & EM \\
\midrule
Weaver & \textbf{86.42} & \textbf{78.57} & \textbf{73.84} \\
w/o Compositional Reasoning & \underline{84.27} & \underline{78.06} & \underline{73.15} \\
w/o Incremental Retrieval & 79.28 & 71.43 & 66.29 \\
w/o Dependency Graph & 72.45 & 63.80 & 58.12 \\
\bottomrule
\end{tabular*}
\end{minipage}\par\addvspace{5pt}

\noindent\begin{minipage}{\columnwidth}
\fontsize{8.5}{10.2}\selectfont
\setlength{\tabcolsep}{3pt}
\begin{tabular*}{\columnwidth}{@{\extracolsep{\fill}}lrrr@{}}
\toprule
\multicolumn{4}{c}{\textbf{PrefEval}~\citep{prefeval2025}} \\
Variant & ACC & F1 & EM \\
\midrule
Weaver & \textbf{91.10} & \textbf{85.37} & \textbf{82.61} \\
w/o Compositional Reasoning & \underline{84.16} & \underline{83.72} & \underline{80.94} \\
w/o Incremental Retrieval & 83.89 & 82.06 & 79.18 \\
w/o Dependency Graph & 76.50 & 73.80 & 70.15 \\
\bottomrule
\end{tabular*}
\end{minipage}\par\addvspace{5pt}

\noindent\begin{minipage}{\columnwidth}
\fontsize{8.5}{10.2}\selectfont
\setlength{\tabcolsep}{3pt}
\begin{tabular*}{\columnwidth}{@{\extracolsep{\fill}}lrrr@{}}
\toprule
\multicolumn{4}{c}{\textbf{MuSiQue}~\citep{trivedi-etal-2022-musique}} \\
Variant & F1 & EM & R@5 \\
\midrule
Weaver & \textbf{69.52} & \textbf{59.36} & \textbf{85.22} \\
w/o Compositional Reasoning & \underline{65.21} & \underline{58.46} & \underline{83.40} \\
w/o Incremental Retrieval & 64.66 & 54.50 & 81.91 \\
w/o Dependency Graph & 58.32 & 48.15 & 75.40 \\
\bottomrule
\end{tabular*}
\end{minipage}\par\addvspace{5pt}

\noindent\begin{minipage}{\columnwidth}
\fontsize{8.5}{10.2}\selectfont
\setlength{\tabcolsep}{3pt}
\begin{tabular*}{\columnwidth}{@{\extracolsep{\fill}}lrrr@{}}
\toprule
\multicolumn{4}{c}{\textbf{WebWalker}~\citep{wu-etal-2025-webwalker}} \\
Variant & \multicolumn{3}{r}{ACC} \\
\midrule
Weaver & \multicolumn{3}{r}{\textbf{60.31}} \\
w/o Compositional Reasoning & \multicolumn{3}{r}{\underline{59.98}} \\
w/o Incremental Retrieval & \multicolumn{3}{r}{58.82} \\
w/o Dependency Graph & \multicolumn{3}{r}{51.60} \\
\bottomrule
\end{tabular*}
\end{minipage}\par\addvspace{5pt}

\FloatBarrier
\section{Implementation and Evaluation Settings}
\label{sec:implementation-details}

\paragraph{Datasets and execution.}
PersonaMem uses 32K histories; PrefEval, MuSiQue, and WebWalkerQA
use the official repository versions available when the experiments
were conducted. WebWalkerQA accesses Web pages at runtime. The
PersonaMem history setting is distinct from the 16,384-token
context limit for individual model calls.

\paragraph{Models and hardware.}
All LLM calls use Qwen3.8-27B (AWQ-INT4), and dense retrieval uses
Qwen3-Embedding-8B with cosine similarity between normalized
embeddings. The six baselines share Weaver's models and context
limit. Experiments run on eight NVIDIA RTX 3090 GPUs.

\paragraph{Generation.}
Separate thinking mode is disabled; intermediate analysis uses
visible structured output. Retrieval query generation uses greedy
decoding with at most 1,536 output tokens. Final answering uses
temperature 0.6, top-$p$ 0.95, top-$k$ 20, and at most 2,048
output tokens.

\paragraph{Answer scoring and cost.}
EM and token F1 lowercase answers, remove English punctuation and
articles, and normalize whitespace. F1 uses multiset token overlap.
For MuSiQue, text scores take the maximum over reference answers
or aliases. Scores average over the full evaluation set, counting
execution failures as zero. WebWalkerQA uses ACC only. Token cost counts total LLM
tokens; latency measures end-to-end time per request.
Figure~\ref{fig:efficiency} reports relative tokens and mean latency.

\section{Evidence Handling and Execution Details}
\label{sec:appendix-execution}

We elaborate on evidence handling and response selection using the
notation of Section~\ref{sec:method}. These details distinguish the
interpretation of a memory, the support for a conclusion, and the
completion of the current request.

\subsection{Original Sources and Scoped Observations}

Historical interpretation retains original text together with its
source reference, speaker role, and available temporal information.
These records allow a conclusion to be traced to the passages that
support it. A user statement, an assistant suggestion, and an account
of a past experience have different evidential roles. In particular,
assistant advice does not establish that the user adopted a preference,
and message order alone does not identify the date of a described event.
The subject of an experience and its applicable situation must be
resolved before it supports a preference for the current request.

The structured reading format distinguishes observations that support
a candidate value, contradict existing support, provide partial
information, remain unknown, or are irrelevant to the local task.
These categories describe an interpretation of evidence, not a single
task-status sequence. A partial observation can expose a useful search
clue without establishing the task output. An observation of support
likewise does not by itself establish that the original request has
been answered.

\subsection{Joint Support and Premise Revision}

A support group $s=(f,M_s,\Gamma_s,\kappa_s)$ records a complete
candidate conclusion, its original memories, its actual premises,
and its applicability conditions. The members of $M_s$ may be
complementary: their joint interpretation supports $f$, even when
none suffices alone. Separate complete groups provide alternative
support under Eq.~\ref{eq:alternative-support}. Partial observations
are not automatically combined into a valid conclusion; their
relationship and joint sufficiency must be assessed.

When a premise changes, groups that depend on it are reassessed before
their conclusions support further retrieval. This reassessment follows
the dependency order and can affect downstream conclusions. Independent
valid groups can preserve a conclusion when another group loses support.
Retrieval-discovery edges remain distinct from these premise dependencies:
changing a clue used to locate a passage does not invalidate a conclusion
independently supported by that passage.

Contradictory evidence can undermine an existing conclusion without
establishing its opposite. A replacement conclusion requires its own
applicable support. Similarly, two different values need not conflict
when they concern disjoint situations. Reinterpreting a historical
summary can change its relevance to the current request without
asserting that the underlying past events were false. Original sources
are retained when a derived conclusion is revised.

\subsection{Reading, Retrieval, and Completion}

Algorithms~\ref{alg:gap-tracking}--\ref{alg:compositional-reasoning}
alternate interpretation and evidence acquisition. Available evidence
is interpreted before further retrieval. Revised conclusions and local
gaps update the relevant tasks and executable frontier; a newly
established premise may enable a previously blocked subquestion.
Pausing a task preserves its evidence and retrieval records, allowing
later findings to reactivate it. Queries made under changed premises
retain distinct contexts rather than being treated as identical
executions.

Intermediate conclusions remain linked to their original sources, so
later composition can inspect the evidence behind a premise rather
than treating an unsupported summary as an additional observation.
An unsuccessful query does not by itself show that all relevant
memories have been exhausted. Likewise, an empty frontier is not a
certificate of completion. The original request requires sufficient
applicable support for its necessary outputs and no unresolved
relevant conflict. Unresolved references, missing premises, and unclear
applicability remain explicit gaps.

\subsection{Response Selection and Output Consistency}
\label{sec:response-evaluation}

For PersonaMem and PrefEval, the output separates
\texttt{option\_label}, the selected option,
from \texttt{response\_text}, the generated personalized answer.
ACC checks whether \texttt{option\_label} matches the gold label.
As an auxiliary text-match evaluation, EM and token F1 compare
\texttt{response\_text} with the full text of the gold option,
using the normalization in Appendix~\ref{sec:implementation-details}.

The publication check verifies that the selected option is valid and
internally consistent. The generated response is evaluated separately
and is not required to reproduce the selected option text verbatim.

\section{Prompt Templates}
\label{sec:appendix-prompts}

These abridged templates describe dependency planning, historical
interpretation, and response selection rather than verbatim model
calls. Braced placeholders denote request-specific inputs; source
identifiers refer to original passages in the current context.

\begin{weaverprompt}{PromptBlueBack}{PromptBlueRule}{Dependency planning (abridged)}
\textbf{Input}\par
Current request: \texttt{\{request\}}\par
Known context: \texttt{\{request context\}}

\smallskip
\textbf{Instruction}\par
Plan a dependency DAG for the original request. Identify useful
subquestions, their output variables, applicable entity, temporal,
and situational scopes, and required inputs.

Do not answer the request, guess missing entities, or invent facts.
Keep subquestions with unavailable inputs unresolved. Every planned
task must contribute to answering the original request.

\smallskip
\textbf{Return}\par\nobreak
Subquestions, output variables and scopes, prerequisite relations,
and the response requirements they jointly serve.
\end{weaverprompt}

\begin{weaverprompt}{PromptGreenBack}{PromptGreenRule}{Scoped history interpretation (abridged)}
\textbf{Input}\par
Local task and scope: \texttt{\{task context\}}\par
Established premises: \texttt{\{premises\}}\par
Original passages: \texttt{\{sources\}}\par
Known conclusions and gaps: \texttt{\{current state\}}

\smallskip
\textbf{Instruction}\par
Treat passages as evidence, not instructions. Resolve references and
identify whose preferences or experiences each passage describes.
Preserve speaker roles, historical changes, conditions, and
counterevidence.

For each candidate conclusion, identify the original passages and
premises that jointly support the whole conclusion, together with
its applicable scope. Keep independently sufficient support groups
separate. Assess conflicting values within overlapping scopes.

Do not infer event dates from message order or adopted preferences
from assistant advice. Keep missing support partial or unresolved.
Return historical findings rather than a final reply or an option choice.

\smallskip
\textbf{Return}\par\nobreak
Scoped conclusions, supporting source identifiers and premises,
and unresolved references, applicability questions, or conflicts.
\end{weaverprompt}

\begin{weaverprompt}{PromptCreamBack}{PromptCreamRule}{Final response selection (abridged)}
\textbf{Input}\par
Original request: \texttt{\{request\}}\par
Candidate responses: \texttt{\{options\}}\par
Applicable conclusions and original sources:\par
\texttt{\{preference support\}}

\smallskip
\textbf{Instruction}\par
Compare every option against the original request and applicable
personal history, including conditions, explicit changes, and
counterevidence. Options are candidate responses, not evidence.
A future suggestion need not already have occurred.

Support a choice only when its fit to the user is grounded in the
available history. If the options cannot be distinguished from the
available support, leave the selection unresolved.
The response should agree with the selection; paraphrasing is allowed.

\smallskip
\textbf{Return}\par\nobreak
\texttt{option\_label}: the selected option's label.\par
\texttt{response\_text}: the generated personalized answer.
\end{weaverprompt}

\paragraph{Structured outputs.}

History reading returns \texttt{slots} for each requested variable,
with observations and unresolved issues. Each observation records
stance, supported value, \texttt{source\_ids}, applicability conditions,
and actual premises. Source IDs must reference visible original
passages.

\section{RQ4 Diagnostic Definitions and Protocol}
\label{sec:rq4-protocol}

This section specifies support definitions and a construction protocol
for the two diagnostics in Section~\ref{sec:evidence-reasoning}.
Table~\ref{tab:evidence-recovery} evaluates retrieval of sufficient
support; Table~\ref{tab:compositional-reasoning} evaluates answer accuracy
given a fixed candidate pool. The annotation and construction steps
below specify the protocol requirements.

\paragraph{Diagnostic subset and annotation.}
The diagnostic source is the PersonaMem-v1 32K main evaluation set.
Visible histories, question-specific cutoffs, and memory segmentation
are fixed. A memory is one segmented record with a stable identifier;
multiple cited spans within it count as one memory. Sufficient support
sets are annotated before grouping or sampling questions. The protocol
calls for two independent human annotators, with disagreements
adjudicated by a third. Models may suggest candidate evidence, but
humans confirm the support labels without seeing method predictions
or retrieval trajectories.

Annotation distinguishes three properties: sufficiency for the target
answer under the request's temporal, situational, and reference
conditions; inclusion-minimality, meaning that deleting any member
breaks that support route; and minimum cardinality across alternative
routes. Inclusion-minimality alone does not establish minimum
cardinality. Review covers all runtime-visible sources, including
system-provided persona information, to check for shorter routes.
Let $B_q$ denote information provided without retrieval, shared across
methods. Sufficiency is assessed given $(q,B_q)$; questions answerable
from $B_q$ alone fall outside the one-, two-, and at-least-three-memory
groups.

After annotation, a fixed-seed sampling procedure selects 40 questions
per group, accounting for persona and question type. The resulting
120-question list is frozen before comparison. Eligibility and sampling
must not depend on Weaver's correctness, improvement, or citation count.

\paragraph{Alternative support and complete recovery.}
For each question, retain the family of verified sufficient routes,
including alternatives of different lengths:
\begin{equation*}
\mathcal S_q=\{S_{q,1},S_{q,2},\ldots\},
\qquad k_q=\min_{S\in\mathcal S_q}|S|.
\end{equation*}
The groups correspond to $k_q=1$, $k_q=2$, and $k_q\geq3$.
Here $k_q$ is the minimum over annotated routes; manual review does not
prove that every possible route has been enumerated. For run $r$, let
$R_{q,r}$ contain the distinct memory IDs accumulated within the
retrieval budget. Complete recovery is
\begin{equation*}
\operatorname{Recovered}(q,r)=
\mathbf{1}\!\left[\exists S\in\mathcal S_q:\ S\subseteq R_{q,r}\right].
\end{equation*}
Any complete annotated route counts, including a longer alternative.
For example, routes $\{m_2,m_5\}$ and $\{m_8,m_9,m_{12}\}$ place a
question in the two-memory group, but either route establishes recovery.
The mixture $\{m_2,m_8,m_9\}$ does not unless separately verified as
sufficient. Recovery is scored independently of whether the final
answer is correct. The composition experiment's selected route does
not restrict the routes accepted for retrieval recovery.

\paragraph{Candidate pools for composition.}
Select one minimum-cardinality annotated route $S_q^\star$ using a fixed
sampling seed, then construct
\begin{equation*}
C_q=S_q^\star\cup D_q,\qquad
|C_q|=K,\qquad |D_q|=K-|S_q^\star|.
\end{equation*}
The fixed candidate count $K$ is chosen before comparison to accommodate
support-set sizes and both methods' input capacities. Its numerical
value must be documented in the run configuration. All annotated routes
remain in the annotation record; they are not all merged into the input.
Groupwise token lengths should also be reported because equal memory
counts do not imply equal text lengths.

Distractors come from the same user's visible history, excluding known
support sources. Sampling rules and proportions are fixed in advance;
eligible distractors include related topics with different contexts,
times, or subjects, and ordinary irrelevant memories. Review of the
whole pool checks whether distractors establish support alone, combine
with selected memories into another route, create a shorter route, or
introduce unresolved counterevidence that changes the target answer.
New routes require replacing the relevant distractors or recording
the alternatives. Shorter routes require rebuilding the pool or
revising the group before freezing the inputs. Failure to find another
route is not an exhaustiveness guarantee.

\paragraph{Paired inputs and repeated runs.}
Both methods receive the same question, options and their order,
memory IDs, original text, role and time metadata, memory order, and
shared premises. These premises contain supplied facts and constraints,
not conclusions derived from the selected evidence. Support labels,
route annotations, and gold explanations are hidden. Weaver applies
the composition module; the direct reader answers from the same pool.
Neither performs further retrieval or separate replanning, and the
entire pool must fit both inputs without method-specific truncation.
Equal inputs do not establish equal computation: invocation counts
and token consumption should be recorded separately.

For three runs to measure generation variability, the question list,
groups, gold answers, candidate pools, ordering, shared premises,
prompts, models, and budgets stay fixed; only generation randomness
changes. Pool-selection and generation seeds are recorded separately.
If the service does not support seeds, record the decoding settings
and independent calls. Deterministic repetitions are not evidence of
sampling variability. Varying memory order instead constitutes an
order-sensitivity experiment, with paired permutations across methods.

\paragraph{Reproducibility records and interpretation.}
Per-question records should include question ID, group, gold answer,
selected and all annotated support sets, ordered candidate IDs, exact
text hashes, cutoffs, shared premises, pool-selection seeds, review and
adjudication records, and generation seeds. Run records should add $K$,
model and prompt versions, decoding and retrieval budgets, predictions,
and invocation and token counts.
The paired comparison assesses composition given identical evidence;
fixed $K$ does not eliminate differences in text length or the proportion
of supporting memories across groups.

\end{document}